\documentstyle[preprint,version2,aps]{revtex}
\begin{document}
\draft
\preprint{OCIP/C 94-4}
\preprint{UQAM-PHE-94-09}
\preprint{May 1994}
\begin{title}
MEASUREMENT OF THE $WW\gamma$ VERTEX THROUGH \\
SINGLE PHOTON PRODUCTION AT $e^+e^-$ COLLIDERS
\end{title}
\author{Gilles Couture\footnote{e-mail: couture@osiris.phy.uqam.ca}}
\begin{instit}
D\'epartement de  Physique, Universit\'e du Qu\'ebec \`a Montr\'eal \\
C.P. 8888, Succ. Centre-Ville, Montr\'eal, Qu\'ebec, Canada, H3C 3P8
\end{instit}
\moreauthors{Stephen Godfrey\footnote{e-mail: godfrey@physics.carleton.ca}}
\begin{instit}
Ottawa-Carleton Institute for Physics \\
Department of Physics, Carleton University, Ottawa CANADA, K1S 5B6
\end{instit}
\begin{abstract}
We perform a detailed study of the process
$e^+e^-\to \gamma\nu_l\bar\nu_l$ and its sensitivity to anomalous gauge
boson couplings of the $\gamma WW$ vertex.
We concentrate on
LEP II energies, $\sqrt{s}=200$ GeV, and energies appropriate to the
proposed Next Linear Collider (NLC) high energy $e^+e^-$ collider
with center of mass energies $\sqrt{s}=500$
and 1~TeV. At 200 GeV, the process
offers, at best, a consistency check of other processes
being considered at LEP200.  At 500~GeV,
the parameters $\kappa_\gamma$ and $\lambda_\gamma$
can be measured to about $\pm 0.05$ and $\pm 0.1$ respectively at 95\% C.L.
while at 1 TeV, they can be measured to about $\pm 0.02$.  At the
high luminosities anticipated at high energy linear colliders precision
measurements are likely to be limited by systematic rather than statistical
errors.
\end{abstract}

\narrowtext
\section{INTRODUCTION}
\label{sec:intro}

The major preoccupation of particle physics is the search for physics
beyond the standard model or equivalently, for
deviations from standard model predictions.  To this end,
measurements at the CERN LEP-100 $e^+e^-$ collider
and the SLAC SLC $e^+e^-$ collider\cite{lep} have provided
stringent tests \cite{hollik,bigfit} of the standard model of the
electroweak interactions
\cite{sm}.  However, it is mainly  the fermion-gauge boson
couplings that have been tested and the gauge sector of the
standard model remains poorly constrained.
A stringent test of the gauge structure of the standard model is provided by
the
tri-linear gauge vertices (TGV's); the $\gamma WW$ and $ZWW$ vertices. Within
the
standard model, these couplings are uniquely determined
by $SU(2)_L \times U(1)$ gauge symmetry so that a precise measurement
of the vertex poses a severe test of the gauge structure of the theory.
If these couplings were observed to have different values than
their standard model values, it would indicate the need for physics beyond
the standard model.

The study of the trilinear
gauge boson couplings by studying $W$ pair production
is one of the primary motivations for the LEP200 upgrade
\cite{hagiwara87,lep200,kane89} with a precision of 30-40\% is expected
from cross section and $W$ angular distribution measurements.
In the far future there is growing interest in the physics that can be done
at high energy $e^+e^-$ colliders with $\sqrt{s}=500$ GeV or $\sqrt{s}=1$ TeV,
referred to as  the Next Linear
Collider (NLC), the Japan Linear Collider (JLC)  or the CERN Linear Collider
(CLIC) \cite{nlcphysics,miyamoto,nlc,JLC,NLC2,CLIC}.
Various options are being studied including $e\gamma$ and $\gamma\gamma$
collisions
where the energetic photons are obtained  by backscattering a laser on
one of the incident leptons.
Measurements at these colliders are very sensitive to anomalous
couplings with $e\gamma$ and $\gamma\gamma$ collisions
putting some of the more stringent bounds on anomalous $WW\gamma$
couplings \cite{yehudai,couture,choi91}.

A problem common to many processes used to study TGV's is that they involve
both the $WW\gamma$ and $WWZ$ vertices making it difficult to disentangle
the contributions. In a previous paper
we presented a detailed study of the process $e^+e^- \to
\nu_l\bar\nu_l \mu^+ \mu^-$ motivated by our interest in isolating
the $WWZ$ and $WW\gamma$ vertices by appropriate kinematic cuts on
the invariant mass of the $\mu^+\mu^-$ \cite{eptoz}.
Included in this
final state are contributions from the underlying process
$e^+e^- \to \nu_l \bar\nu_l \gamma^* \to \nu_l\bar\nu_l \mu^+ \mu^-$ which
shows up most dramatically when $M_{\mu^+\mu^-}\to 0$.  However, because
of the muons' masses it does not quite isolate the process we are
interested in.  In this paper we take the obvious limit and study the
sensitivity of the process $e^+e^- \to \gamma \nu_l\bar\nu_l$ to anomalous
$WW\gamma$ couplings\cite{miyamoto,eptog}.
This process has also been used as a means of
counting the number of light neutrino species\cite{neutcount}.

To parametrize the $WW\gamma$ vertex we use the most general
parametrization possible that respects Lorentz invariance,
electromagnetic gauge invariance and $CP$ invariance
\cite{hagiwara87,gaemers79,miscvertex} since it has become the standard
parametrization used in phenomenology and therefore makes the comparison
of the sensitivity of different measurements to the TGV's straightforward.
We do not
consider CP violating operators in this paper as they  are tightly
constrained by measurement of the neutron
electron dipole moment which constrains the two CP violating parameters to
$|\tilde{\kappa}_\gamma|, |\tilde{\lambda}_\gamma|<
{\cal O} (10^{-4})$ \cite{cp}.
Therefore the $WW\gamma$ vertex has two free independent
parameters, $\kappa_\gamma$ and $\lambda_\gamma$ and
is given by \cite{hagiwara87,gaemers79}:
\begin{equation}
{\cal L}_{WW\gamma} =  - ie \left\{ { (W^\dagger_{\mu\nu}W^\mu A^\nu -
W^\dagger_\mu A_\nu W^{\mu\nu} )
+ \kappa_\gamma W^\dagger_\mu W_\nu F^{\mu\nu}
- {{\lambda_\gamma}\over{M_W^2}} W^\dagger_{\lambda\mu}W^\mu_\nu F^{\nu\lambda}
}\right\}
\end{equation}
where $A^\mu$ and $W^\mu$ represents  the photon and $W^-$
fields respectively,  $W_{\mu\nu}=\partial_\mu W_\nu-\partial_\nu W_\mu$ and
$F_{\mu\nu}=\partial_\mu A_\nu-\partial_\nu A_\mu$ where $A$ is the
photon and $M_W$ is the $W$ boson mass.
Higher dimension operators would correspond to momentum dependence in the form
factors which we ignore.
At tree level the standard model requires $\kappa_\gamma=1$ and
$\lambda_\gamma=0$.
Note that the presence of the W-boson mass factor in the
$\lambda_V$ term is {\it ad hoc} and one could argue that the scale $\Lambda$
of new physics would be more
appropriate. We will conform to the usual parametrization and will not address
this issue any further.

We studied the sensitivity of this process at TRISTAN and LEP/SLC
energies where there exists data\cite{tristan,lepnc}
that we could in principle use to bound
the $WW\gamma$ couplings.  However, we found that the process was
insufficiently sensitive at these energies to put meaningful bounds on
the $WW\gamma$ coupling with the integrated luminosities already
accumulated or expected in the foreseeable future.
We therefore start with
$\sqrt{s}=200$ GeV appropriate to LEP200 since this machine will be operational
in the
relatively near future \cite{lep200}. We then turn to the proposed JLC/NLC/CLIC
$e^+e^-$
colliders with possible center of mass energies of $\sqrt{s}=500$
GeV and  1 TeV\cite{nlc,JLC,NLC2,CLIC}.
We do not include any beamsstrahlung radiation effects
in our calculation \cite{beam}. These effects are
very much machine dependant (beam intensity, bunch geometry, etc \dots)
and known to be
negligible at 200 GeV, and small at 500 GeV. However, although
they can be quite important at 1000 GeV,
there has been progress in strategies to minimize the
effects of beamstrahlung radiation.

\section{CALCULATIONS AND RESULTS}
\label{sec:process}
The diagrams contributing to the process $e^+e^-\to \gamma\nu_l\bar\nu_l$
are shown in Fig. 1. The main advantage of this process is that it depends
only on the $WW\gamma$ vertex. In addition, our signal (fig. 1(a)) should
increase with energy, for two reasons: it is a t-channel process and should not
decrease as fast with energy as the other contributions to the total
process, especially when suitable kinematic cut are imposed to eliminate
the on-shell $Z$ contribution. Also,
anomalous couplings, in general, become more important at higher energies.

To evaluate the cross-sections and different distributions, we used the CALKUL
helicity amplitude technique \cite{calkul} to obtain expressions for the
matrix elements and
performed the phase space integration using  Monte Carlo
techniques \cite{monte}.
The expressions for the helicity amplitudes are lengthy and unilluminating so
we do not include them here.  The interested reader can obtain them directly
from the authors.
To obtain numerical results we used the values $\alpha=1/128$,
$\sin^2\theta=0.23$, $M_Z=91.187$ GeV, $\Gamma_Z=2.5$ GeV, $M_W=80.2$ GeV, and
$\Gamma_W=2.1$ GeV.

The signal we are studying
is an energetic $\gamma$ plus missing transverse momentum.
The largest potential background is Bhabba scattering with a hard photon
and the electron and positron going down the beam pipe.  This
should not be a serious problem given that we are interested in energetic
photons and the luminosity monitors, based on Bhabba scattering, should
be able to veto any such events.
In order to take into account finite detector acceptance, we
require that the photon be at least 10 degrees away from the beam line
although in practice the cuts we use to enhance the signal are much more
restrictive than this (typically $\sim 30^o$).

In principle we should include QED radiative corrections from soft photon
emission and the backgrounds due to a second photon either that is lost
down the beam pipe or collinear to the hard photon being measured
and therefore unresolved \cite{emrc}.  Although soft photon emission can
reduce the cross section substantially, their inclusion does not
substantially effect the bounds we obtain and therefore our conclusions.
The effects of an unseen second photon turn out to be quite small.  Since
both these contributions depend on details of detector such as energy
resolution and do not alter our conclusions we leave them out but stress
that they must be included in detailed detector Monte Carlo simulations.

The approach we followed was to examine various kinematic distributions,
$d\sigma/d\cos\theta_\gamma$, $d\sigma/dE_\gamma$,
$d\sigma/d{E_{T}}_\gamma$,
and kinematic cuts to find which ones optimized the sensitivity to anomalous
couplings.  In general, the tightest constraints were obtained
by imposing cuts on the
photon energy which eliminated contributions from on-shell $Z^0$ production
(diagrams 1(d) and 1(e)).

\vskip 0.3cm

\begin{center}
A. $\sqrt{s}=200$ GeV
\end{center}

\vskip 0.2cm

In Fig. 2 we show the $E_\gamma$ distribution for $\sqrt{s}=200$~GeV for
several values of $\kappa_\gamma$ and $\lambda_\gamma$.  The most
prominent feature of the distribution comes from the contribution
of the on-shell $Z^0$ (diagrams 1(d) and 1(e)).  From Fig. 2 it can be
seen that the regions off the $Z^0$ resonance are most sensitive to
anomalous couplings with the greatest sensitivity in
the region above the $Z^0$ resonance.  This is because that region
probes the largest momentum transfer through the $W$-boson t-channel
propagators.
We found that the tightest constraints could be
obtained by imposing an angular cut of $35^\circ < \theta_\gamma <
145^\circ$
and on the two regions of $E_\gamma$: 25~GeV$<E_\gamma<$65~GeV and
$E_\gamma>$88~GeV.
For 25~GeV$<E_\gamma<$65~GeV the cross section is 0.14~pb
which for an integrated luminosity of 500~pb$^{-1}$ results in about
a 12\% statistical error.  Similarly for $E_\gamma>$88~GeV
$\sigma=0.087$~pb which gives a 50\% statistical error.
Monte Carlo studies of SLD
type detectors give very crude estimates of systematic errors of 5\% for
cross section measurements \cite{barklow}.
Therefore the effects of including a 5\% systematic error are not
particularly important.
The 68\% C.L., 90\% C.L., and 95\% C.L.
bounds that could be obtained with these cuts for $\sqrt{s}=200$~GeV and
integrated luminosity of L=500~pb$^{-1}$ are shown in Fig. 3 with
numerical values when  varying one parameter at a time given in Table I.
It is worth mentioning that the {\it most efficient} energy and angular cuts
will vary slightly with the domain of $\kappa_\gamma$ and $\lambda_\gamma$
being
probed. However, this dependance is mild and the limits will not vary much if
one uses a fixed set of cuts.

An important question for LEP200 is the effect on the {\it physics reach}
of different center of mass energies.  To gauge the change of the
sensitivity to the TGV's for different energies we plot in Fig. 4 the
90\% C.L. for $\sqrt{s}=175$~GeV, 200~GeV, and 230~GeV.  Clearly, the
higher the energy the tighter the constraints that can be obtained.
Roughly speaking, increasing the c.m. energy by 15\% will increase the
sensitivity to anomalous couplings by 30\%.
However, even for the highest possible energies at LEP200 of
$\sqrt{s}=230$~GeV the obtainable bounds are not competitive with the
$W$-pair production process or for that matter with bounds obtained by
the Tevatron experiments via associated $W\gamma$ production,
and will, at best, be a consistency check of other measurements.

\vskip 0.3cm

\begin{center}
B. $\sqrt{s}=500$ GeV
\end{center}

\vskip 0.2cm

We next turn to an ``NLC'' type $e^+e^-$ collider with $\sqrt{s}=500$
GeV.  We consider integrated luminosities of 10 and 50 $fb^{-1}$.  The
photon energy distributions for a several values of $\kappa_\gamma$ and
$\lambda_\gamma$ are shown in Fig. 5.  As before, the values of
$E_\gamma$ most sensitive to anomalous couplings are off the $Z^0$
resonance and  $\kappa_\gamma$ and $\lambda_\gamma$ are sensitive to different
values of $E_\gamma$.  For example, $\kappa_\gamma$ is most sensitive to
$50< E_\gamma < 160$~GeV while $\lambda_\gamma$ is most sensitive to
$177 <E_\gamma < 237$~GeV.
The bounds obtained from these observables are shown in Fig. 6 with
numerical values obtained by varying one parameter at a time given in
Table I.  One could obtain additional information by using the transverse
energy
distribution of the photons with and without
left-handed polarized electrons.
For $50< E_\gamma < 160$~GeV the cross section is $\sigma=0.27$~pb which
for an integrated luminosity of 50~fb$^{-1}$ gives a statistical error of
0.8\%.  Likewise for $177 <E_\gamma < 237$~GeV we obtain
$\sigma=0.024$~pb and
$\delta\sigma^{stat}/\sigma=3$\% and for $E_\gamma>245$~GeV
$\sigma=0.0019$~pb and $\delta\sigma^{stat}/\sigma=10$\%.
Given the small statistical errors including a 5\% systematic errors will
have a considerable effect on the bounds that can be obtained.  For
example for $50< E_\gamma < 160$~GeV where the statistical error is
smallest including a 5\% systematic error reduces the sensitivity by
roughly a factor of 5 while for $177 <E_\gamma < 237$~GeV where the
statistical error is larger it has only a small effect and for
$E_\gamma>245$~GeV where the statistical error is largest the effect of
the systematic error is neglegible.

\vskip 0.3cm

\begin{center}
C. $\sqrt{s}=1$ TeV
\end{center}

\vskip 0.2cm

The final case we consider is a 1 TeV $e^+e^-$ collider.
In Fig. 7 we show the photon energy distribution for several values of
$\kappa_\gamma$ and $\lambda_\gamma$.  As in the 500~GeV case we find
that $\kappa_\gamma$ and $\lambda_\gamma$ are sensitive to different
values of $E_\gamma$.
For the photon energy range $60 < E_\gamma < 220$~GeV  $\sigma=0.39$~pb
which for an integrated luminosity of 200~fb$^{-1}$ gives
$\delta\sigma^{stat}/\sigma=0.3$\% while for $245 < E_\gamma < 490$~GeV
$\sigma=0.05$~pb $\delta\sigma^{stat}/\sigma=1$\%. Including
a 5\% systematic error weakens the bounds on the TGV's considerably,
even more so at $\sqrt{s}=1$~TeV than at 500~GeV.

In Fig. 8 we show the bounds that could be
obtained using $60 < E_\gamma < 220$~GeV and
$245 < E_\gamma < 490$~GeV and give some numerical values in Table I.
The bounds obtainable on $\kappa_\gamma$ are
down to the percent level necessary to probe radiative corrections to the
gauge boson vertices.  Although the bounds on $\lambda_\gamma$ are also
down to this level radiative corrections are about an order of magnitude
smaller so that it is unlikely that deviations from the tree-level value
could be observed unless there was a radical departure from the standard
model by, for example, compositeness.

\section{CONCLUSIONS}

We have examined the usefulness of the process $e^+e^-\to \gamma\nu_l\bar\nu_l$
for measuring the $\gamma W^+W^-$ vertex.
The sensitivity of this process to anomalous coupling is
greatly enhanced (by a factor of 5-8 generally) by eliminating the
on-shell $Z$
and by splitting the energy domain of the photon into more
than one bin.
The process turned out to be too insensitive at TRISTAN and LEP-100/SLC
energies to obtain bounds competitive with recent results from the
Tevatron.  In fact, the same is also true for LEP-200 energies which at
best will offer a consistency check
of bounds extracted from $W$-pair production at LEP-200 and the Tevatron.
At higher energy
$e^+e^-$ colliders, this process
can lead to very stringent bounds, precise enough to
test the TGV's at the level of radiative corrections.
We used the
high luminosities planned for at the high energy $e^+e^-$ colliders to
estimate statistical errors.  When we included reasonable estimates
of systematic errors we found that the limiting factor in high precision
measurements will likely be systematic errors not statistical errors.
However, we based our analysis on optimizing cuts on  $E_\gamma$ and
$E_{T\gamma}$ and it is likely that performing a more detailed maximum
likelihood fit of real data would make fuller use of the information in
the data thereby obtaining tighter bounds.
The challenge will
be to reduce the systematic errors and one should be very
careful with respect to the conclusions one makes by considering only
statistical errors.

\acknowledgments

This research was supported in part by the Natural Sciences and
Engineering Research Council of Canada and Les Fonds FCAR du
Quebec. S.G. thanks Dean Karlen for helpful conversations.

\figure{The Feynman diagrams contributing to the process
$e^+e^- \to \gamma \nu \bar{\nu}$
\label{diagrams}}

\figure{The photon energy distribution, $d\sigma/dE_\gamma$ at
$\sqrt{s}=200$~GeV.  The
solid line is for standard model values of $\kappa_\gamma$ and
$\lambda_\gamma$,
the long dashed line is for $\kappa_\gamma=-3$ and $\lambda_\gamma=0$;
the dotted line is for $\kappa_\gamma=5$ and $\lambda_\gamma=0$;
the dot-dashed line is for $\kappa_\gamma=1$ and $\lambda_\gamma=-4$;
and the dot-dot-dashed line is for $\kappa_\gamma=1.0$ and
$\lambda_\gamma=4$.}

\figure{Sensitivities of the TGV's to anomalous couplings
for $\sqrt{s}=200$~GeV and L=500~pb$^{-1}$.  The horizontal curves are
based on $25<E_\gamma < 65$~GeV and the vertically aligned oblongs are
based on $E_{T\gamma}>84$~GeV.  In both cases the solid curves represent
68\% CL limits, the dashed curves 90\% CL limits and the dot-dashed
curves 95\% CL limits.}

\figure{Sensitivities of the TGV's to anomalous couplings at 90\% CL for
$\sqrt{s}=230$~GeV (solid curves),
$\sqrt{s}=200$~GeV (dashed curves), and
$\sqrt{s}=175$~GeV (dot-dashed curves) all for 500~pb$^{-1}$ integrated
luminosity.
The horizontal curves are based on
$25<E_\gamma<80$~GeV for $\sqrt{s}=230$~GeV, and
$25<E_\gamma<65$~GeV for $\sqrt{s}=200$~GeV,
while the vertical oblongs are based on
$E_\gamma>104$~GeV for $\sqrt{s}=230$~GeV,
$E_\gamma>88$~GeV for $\sqrt{s}=200$~GeV, and
$E_\gamma>72$~GeV for $\sqrt{s}=175$~GeV.}

\figure{The photon energy distribution, $d\sigma/dE_\gamma$ for
$\sqrt{s}=500$~GeV.  The
solid line is for standard model values of $\kappa_\gamma$ and
$\lambda_\gamma$,
the long dashed line is for $\kappa_\gamma=0.0$ and $\lambda_\gamma=0$;
the dotted line is for $\kappa_\gamma=2.0$ and $\lambda_\gamma=0$;
the dot-dashed line is for $\kappa_\gamma=1$ and $\lambda_\gamma=-1.0$;
and the dot-dot-dashed line is for $\kappa_\gamma=1.0$ and
$\lambda_\gamma=1.0$.}

\figure{Sensitivities of the TGV's to anomalous couplings
for $\sqrt{s}=500$~GeV and L=50~fb$^{-1}$.
The horizontal curves are based on $50<E_\gamma< 160$~GeV;
the vertical curves are based on $E_\gamma>245$~GeV; and the
upside-down
U shaped curves are based on $177<E_\gamma<237$~GeV.
In all cases the solid lines are 68\% C.L.
and the dot-dashed curves are 95\% C.L..}

\figure{The photon energy distribution, $d\sigma/dE_\gamma$ for
$\sqrt{s}=1000$~GeV.  The
solid line is for standard model values of $\kappa_\gamma$ and
$\lambda_\gamma$,
the long dashed line is for $\kappa_\gamma=0.2$ and $\lambda_\gamma=0$;
the dotted line is for $\kappa_\gamma=1.8$ and $\lambda_\gamma=0$;
the dot-dashed line is for $\kappa_\gamma=1$ and $\lambda_\gamma=-0.2$;
and the dot-dot-dashed line is for $\kappa_\gamma=1.0$ and
$\lambda_\gamma=0.2$.}

\figure{Sensitivities of the TGV's to anomalous couplings
for $\sqrt{s}=1$~TeV and L=200~fb$^{-1}$.
The horizontal curves are based on $60<E_\gamma< 220$~GeV;
and the inverted U shaped curves are based on $245<E_{\gamma}<490$~GeV.
In both cases the solid lines are 68\% C.L., the dashed lines are 90\% C.L.,
and the dot-dashed curves are 95\% C.L..}

\newpage
\begin{table}
\caption{Sensitivities to $\kappa_\gamma$ and $\lambda_\gamma$ at 95\% C.L.
from the process $e^+e^- \to \gamma \nu \bar{\nu}$
at a 200~GeV, 500~GeV, and 1~TeV $e^+e^-$ colliders.
The statistical error is based on the specified integrated luminosity.
The entry $---$ denotes a bound too weak to be relevant.}
\begin{tabular}{llll}
\multicolumn{4}{c}{$\sqrt{s}=200$~GeV} \\
\tableline
	Observable &
	& $\delta^{stat}_1$(L=250~pb$^{-1}$)
	& $\delta^{stat}_2$(L=500~pb$^{-1}$) \\
\tableline
$25< E_\gamma < 65$~GeV
	& $\delta\kappa_\gamma$
	& ${ ---\atop -2.5}$ & ${ ---\atop -1.9}$  \\
$ E_\gamma >88$~GeV
	& $\delta\kappa_\gamma$
	& ${+ 2.3\atop -3.2}$ & ${+ 1.9\atop -2.6}$ \\
	& $\delta\lambda_\gamma$
	& ${+ 2.9\atop -2.4}$ & ${+ 2.5\atop -2.0}$ \\
$ E_{T\gamma} >84$~GeV
	& $\delta\kappa_\gamma$
	& ${+ 2.9\atop -3.3}$ & ${+ 2.5\atop -2.8}$ \\
	& $\delta\lambda_\gamma$
	& ${+ 2.8\atop -2.4}$ & ${+ 2.4\atop -1.9}$ \\
\tableline
\tableline
\multicolumn{4}{c}{$\sqrt{s}=500$~GeV} \\
\tableline
	Observable &
	& $\delta^{stat}_1$(L=10~fb$^{-1}$)
	& $\delta^{stat}_2$(L=50~fb$^{-1}$)\\
\tableline
$50< E_\gamma < 160$~GeV
	& $\delta\kappa_\gamma$
	& ${+ .13\atop -.12}$ & ${+ .055\atop -.055}$ \\
$177< E_\gamma < 237$~GeV
	& $\delta\kappa_\gamma$
	& ${+ .16\atop ---}$ & ${+ .08\atop -.10}$ \\
	& $\delta\lambda_\gamma$
	& ${+ .16\atop -.10}$ & ${+ .12\atop -.06}$ \\
$ E_\gamma >245$~GeV
	& $\delta\lambda_\gamma$
	& ${+ .17\atop -.23}$ & ${+ .10\atop -.17}$ \\
\tableline
\tableline
\multicolumn{4}{c}{$\sqrt{s}=1000$~GeV} \\
\tableline
	Observable &
	& $\delta^{stat}_1$(L=50~fb$^{-1}$)
	& $\delta^{stat}_2$(L=200~fb$^{-1}$) \\
\tableline
$60< E_\gamma < 220$~GeV
	& $\delta\kappa_\gamma$
	& ${+ .037\atop -.036}$ & ${+ .018\atop -.018}$  \\
$245< E_\gamma < 490$~GeV
	& $\delta\lambda_\gamma$
	& ${+ .039\atop -.015}$ & ${+ .034\atop -.009}$ \\
\end{tabular}
\end{table}

\end{document}